\input{switches}
\documentclass[sigplan,%
  \ifbool{arxiv}{nonacm}{\ifbool{final}{}{review}}
]{acmart}
\usepackage{subfiles}
\usepackage[\xcoloropt]{xcolor}
\usepackage{cleveref}
\usepackage{braket}
\usepackage{tikz}
\usetikzlibrary{matrix,decorations,arrows,calc,trees}
\usepackage{tcolorbox}
\tcbuselibrary{listings,xparse}
\usepackage{enumitem}

\Crefformat{enumi}{(#2#1#3)}

\definecolor{ltblue}{rgb}{0,0.4,0.4}
\definecolor{dkblue}{rgb}{0,0.1,0.6}
\definecolor{dkgreen}{rgb}{0,0.35,0}
\definecolor{dkviolet}{rgb}{0.3,0,0.5}
\definecolor{dkred}{rgb}{0.5,0,0}
\definecolor{comment}{HTML}{444444}
\definecolor{keywd}{HTML}{8D00ED}
\definecolor{types}{HTML}{1F7B2F}
\definecolor{str}{HTML}{4070a0}
\definecolor{code-background}{gray}{0.8}
\definecolor{pragma}{HTML}{372A78}
\definecolor{num}{HTML}{40a070}
\definecolor{symb}{HTML}{000000}

\let\textt=\texttt

\ifbool{arxiv}{%
  \ifbool{arxiv-cached}{%
    \usepackage[frozencache]{minted}%
  }{%
    \usepackage[finalizecache]{minted}%
  }
}{
\usepackage{minted}
}
\ifbool{arxiv}{}{\setminted{style=borland}}

\newminted[code]{haskell}{%
mathescape,linenos,numbersep=5pt,frame=lines,framesep=2mm,%
}
\newminted[repl]{haskell}{%
mathescape,numbersep=5pt,frame=lines,framesep=2mm,%
}
\VerbatimFootnotes
\DeclareTotalTCBox{\hask}{v}{%
tcbox raise base,box align=base,verbatim,colback=lightgray,colframe=gray%
}{\mintinline[fontsize=\SMALL]{haskell}{#1}}
\let\haskinline=\hask

\acmConference[Haskell '21]{Haskell Symposium '21: the 2021 ACM SIGPLAN Symposium on Haskell}{August 26--27, 2021}{Virtual}
\acmBooktitle{Haskell '21: the 2021 ACM SIGPLAN Symposium on Haskell, August 26--27, 2021, Virtual}
\acmPrice{15.00}
\acmISBN{978-1-4503-XXXX-X/18/06}

\begin{document}

\title[Functional Pearl: Witness Me]{Functional Pearl: Witness Me --- Constructive Arguments Must Be Guided with Concrete Witness}

\ifbool{realname}{%
\author{Hiromi Ishii}
\email{h-ishii@math.tsukuba.ac.jp}
\affiliation{%
  \institution{DeepFlow, Inc.}
  \streetaddress{3-16-40}
  \city{Fujimi-shi Tsuruse nishi}
  \state{Saitama prefecture}
  \country{Japan}
  \postcode{354-0026}}
  \renewcommand{\shortauthors}{Hiromi Ishii}
}{}



\begin{abstract}
  Beloved Curry--Howard correspondence tells that types are intuitionistic propositions, and in constructive math, a proof of proposition can be seen as some kind of a construction, or \emph{witness}, conveying the information of the proposition.
  We demonstrate how useful this point of view is as the guiding principle for developing dependently-typed programs.
\end{abstract}

\maketitle

\section{Introduction}
Since Haskell had been given a Promotion~\cite{Yorgey:2012}, using Haskell with dependent types is a joy.
It's not only a joy: it gives us a neat language to express invariants of programs with type-level constraints.
But it also comes with pain: writing correct but maintainable dependently-typed programs in Haskell sometimes is a hard job, as explained by Lindley and McBride~\cite{10.1145/2503778.2503786}.
It is particularly hard when one tries to bridge a gap between types and expressions, to maintain complex type constraints, and so on.

Here, we propose to borrow the wisdom from one of the greatest discovery in Computer Science and Logic: Curry--Howard correspondence.
It tells us that types are propositions and programs are proofs in intuitionistic logic, in a rigorous sense.
According to Brouwer--Heyting--Kolmogorov (BHK) interpretation\footnote{For the precise historical background, we refer readers to the article by Wadler~\cite{Wadler:2015aa}.}, its informal ancestor, a proof of an intuitionistic proposition is some kind of a construction, or \emph{witness}, conveying the information of the proposition.
For example, a proof of $\varphi \land \psi$ is given by a pair of proofs $T: \varphi$ and $S: \psi$; that of $\varphi \to \psi$ is given by a function taking a proof of $P$ and returns that of $Q$, and so on.
The interpretation of our particular interest in this paper is that of disjunction:
\begin{gather*}
  A: \varphi \vee \psi \iff A = \braket{i, B},
\\
\text{where }
i = 0 \text{ and } B: \varphi, \text{ or }
i = 1 \text{ and } B: \psi
\end{gather*}
That is, a witness of disjunction is given by a tuple of the tag recording which case holds and its corresponding witness.
Although BHK interpretation is not as rigorous as Curry--Howard correspondence, its looseness allows us much broader insights, as we shall see in the present paper.

This paper is organised as follows.
\begin{itemize}
  \item In \Cref{sec:gcd}, taking a type-level GCD as an example, we demonstrate how to \emph{demote} closed type-level functions involving pattern-matchings.
  We suggest adding \emph{witnessing} arguments to such type-level functions to make the compiler aware of evaluation paths.
  \item \Cref{sec:disj} demonstrates that we can emulate \emph{disjunctions} of type constraints, provided that constraints in question can be recovered from some statically computable \emph{witnesses}.
  We use a field accessor for a union of extensible records as an example.
  \item \Cref{sec:plugins} shows a practical example of a dependently-typed plugin system type-checkable dynamically at runtime.
  There, we see how the combination of the Deferrable constraint pattern~\cite{Kmett:2020ab} and witness manipulation can be used to achieve this goal.
  We also discuss the design of the witness of type-level equalities.
  \item Finally, we conclude in \Cref{sec:concl}.
\end{itemize}

A complete working implementation is available \ifbool{maingo}{in the \href{https://github.com/konn/demotion-paper/tree/master/demotion-examples}{\texttt{demotion-examples}} directory of the support repository~\cite{demotion-repo}}{as the supplementary tarball}.

\subsection{Preliminaries}
In this paper, we use the standard method of \emph{singletons}~\cite{Eisenberg:2012} to simulate dependent types in Haskell.
Briefly, a singleton type \hask{Sing a} of a type-level value \hask{a} is the unique type that has the same structure as \hask{a}, on which we can pattern-match to retrieve its exact shape.
\hask{Sing a} can be identified with a type \hask{a} but demoted to the expression-level.
In particular, we assume the following API:

\begin{code}
type family Sing :: k -> Type

class Known a where -- SingI in singletons
  sing :: Sing a
withKnown :: Sing a -> (Known a => r) -> r

data SomeSing k where -- SingKind in singletons
  MkSomeSing :: Sing (a :: k) -> SomeSing k
class HasSing k where
  type Demoted k 
  demote :: Sing (a :: k) -> Demoted k
  promote :: Demoted k -> SomeSing k

withPromoted :: HasSing k
  => Demoted k
  -> (forall x. Sing (x :: k) -> r) -> r

type FromJust :: ErrorMessage -> Maybe a -> a
type family FromJust err may where 
  FromJust err 'Nothing = TypeError err
  FromJust _ ('Just a)  = a

type instance Sing = (SNat :: Nat -> Type)
sNat :: KnownNat n => SNat n

withKnownNat :: SNat n -> (KnownNat n => r) -> r
(%+) :: SNat n -> SNat m -> SNat (n + m)
sMod :: SNat n -> SNat m -> SNat (n `Mod` m)
\end{code}

For the detail of singleton-based programming, we refer readers to Eisenberg--Weirich~\cite{Eisenberg:2012} and Lindley--McBride~\cite{10.1145/2503778.2503786}.

We use the following convention:
\begin{enumerate}
\item We prefix singleton types with the capital \hask{S}: e.g.\ \hask{SNat n} is the type of a singleton of \hask{n :: Nat}.
\item For a type-level function we use small \hask{s} as a prefix for singletonised expression-level function: \hask{sMod} is the singletonised version of \hask{Mod}.
\item For operators, we prefix with \haskinline{
\end{enumerate}

\section{Toy Example: Demoting Type-level GCD}
\label{sec:gcd}
Let us begin with a simple example of type-level greatest common divisors (GCDs):

\begin{code}
import GHC.TypeLits

type family GCD n m where
  GCD 0 m = m
  GCD n 0 = n
  GCD n m = GCD (Mod m n) n
\end{code}

So far, so good.

\begin{repl}
>>> :kind! GCD 12 9
GCD 12 9 :: Nat
= 3
\end{repl}

Suppose we want to ``demote'' this definition of \hask{GCD} to expression-level using singletons, that is, to implement the following function \hask{sGCD}:

\begin{code}
sGCD :: SNat n -> SNat m -> SNat (GCD n m)
\end{code}

First, we need to test the equality of type-level naturals.
In the \texttt{base} package, there is a suitable type-classs for it:
\begin{code}
-- Defined in Data.Type.Equality in base
class TestEquality f where
  testEquality :: f a -> f b -> Maybe (a :~: b)

data (:~:) a b where Refl :: a :~: a
\end{code}
Assuming the \hask{TestEquality SNat} instance, one might first attempt to write it as follows:
\begin{code}
sGCD :: SNat n -> SNat m -> SNat (GCD n m)
sGCD sn sm =
  case ( testEquality sn (sNat @0), 
         testEquality sm (sNat @0)) of
    (Just Refl, _) -> sm
    (_, Just Refl) -> sn
    (Nothing, Nothing) -> sGCD (sMod sm sn) sn
\end{code}

The first two cases type-check as expected, but the last case results in the following type error:

\begin{repl}
• Couldn't match type ‘GCD (Mod m n) n’
  with ‘GCD n m’
  Expected type: SNat (GCD n m)
    Actual type: SNat (GCD (Mod m n) n)
  NB: ‘GCD’ is a non-injective type family...
\end{repl}

Why? The definition of \hask{sGCD} seems almost literally the same as type-level \hask{GCD}.
It first match \hask{n} against \hask{0}, then \hask{m} against \hask{0}, and finally fallbacks to \hask{GCD (Mod m n) n}.

Carefully analysing the first two cases, one can realise that there are additional type-level constraints introduced by \hask{Refl} GADT constructor:

\begin{code}
  Refl :: a ~ b => a :~: b
\end{code}

Thus, in the first two cases, the compiler can tell either \hask{n ~ 0} or \hask{m ~ 0}.
Since the \hask{GCD} is defined as a closed type family, the compiler can match clauses in a top-down manner and successfully apply either of the first two clauses of the definition of \hask{GCD}.
In other words, the constructor \hask{Refl} \emph{witnesses} the evaluation path of type-level function \hask{GCD} in the first two cases.

In the last case, however, no additional type-level constraint is available.
Despite humans can still think ``as all the \hask{Refl} clauses failed to match, hence the non-equal clause must apply here'', this intuition is not fully expressed in the type-level constraint!

So we have to give the compiler some \emph{witness} also in the last case.
What kind of a witness is needed here?
Well, we need to teach the compiler which clause was actually used.
In this case, branching is caused by type-level equality: the evaluation path depends on whether \hask{n} or \hask{m} is \hask{0} or not.
First, let us make this intuition clear in the definition of \hask{GCD}:

\begin{code}
import Data.Type.Equality (type (==)) -- from base

type GCD n m = GCD_ (n == 0) (m == 0) n m
type family GCD_ nEq0 mEq0 n m :: Nat where
  GCD_ 'True  _      _ m = m  -- n ~ 0; return m
  GCD_ 'False 'True  n _ = n  -- m ~ 0; return n
  GCD_ 'False 'False n m =    -- Neither; recur!
    GCD_ (Mod m n == 0) 'False (Mod m n) n
\end{code}

Here, we have two type-level functions: newly defined one, \hask{GCD_}, is the main loop implementing Euclidean algorithm, and
\hask{GCD} is redefined to call \hask{GCD_} with the needed information.
Now, \hask{GCD_} takes not only natural numbers but also a type-level \hask{Bool}s \emph{witnessing} equality of \hask{n} and \hask{m} with \hask{0}.
From this, GHC can tell which clause is taken from the first two type-arguments.
As clauses in closed type families can be viewed as a mutually exclusive alternatives, this approach shares the spirit with the constructive BHKs interpretation of $\vee$.

Now that we can give the compiler witnesses as the first two type-arguments of \hask{GCD_}, we are set to implement \hask{sGCD}.
First, we need \emph{demoted} version of type-level \hask{(==)}.
The first attempt might go as follows:
\begin{code}
(%==) :: TestEquality f
      => f a -> f b -> SBool (a == b)
sa %== sb = case testEquality sa sb of
  Just Refl -> STrue
  Nothing -> SFalse
\end{code}
Unfortunately, this doesn't work as expected.
The first error on \hask{STrue} says:

\begin{repl}
• Could not deduce: (a == a) ~ 'True
  from the context: b ~ a
    bound by a pattern with constructor:
               Refl :: forall k (a :: k). a :~: a,
             in a case alternative
    at /.../GCD.hs:33:8-11
  Expected type: SBool (a == b)
    Actual type: SBool 'True
\end{repl}

This is due to the definition of type-level \hask{(==)} in GHC base library:

\begin{code}
type family a == b where
  f a == g b = (f == g) && (a == b)
  a   == a   = 'True
  _   == _   = 'False
\end{code}
As described in the documentation~\cite{GHC-Team:2021aa}, the intuition behind the definition of the first clause is to let the compiler to infer, e.g.\ \hask{Just a == Just b} from \hask{a == b}.

This behaviour is desirable when one treats equalities involving compound types, like \hask{(f a == g b) ~ 'True}.
But when one wants to give a witness of \hask{(a == b) ~ 'True}, we cannot make use of \hask{Refl}.
This is, again, due to the lack of witness of the evaluation path: the compiler cannot determine which clause should be taken to compute \hask{a == b} if \hask{a} and \hask{b} are both opaque variable!

A solution here is just to define another type family, which requires the reflexivity only:
\begin{code}
type family a === b where
  a === a = 'True
  _ === _ = 'False
\end{code}

Although this equality cannot treat equalities between compound types inductively, it suffices for \hask{GCD} case.
We will revisit to a treatment of type-level equality in \Cref{sec:plugins}.
Demoted version of this now gets:

\begin{code}
(%===) :: TestEquality f
       => f a -> f b -> SBool (a === b)
sa %=== sb = case testEquality sa sb of
  Just Refl -> STrue
  Nothing -> SFalse  
\end{code}

Now, the type-error remains on the last clause: \hask{SFalse}.
This is, again, due to the lack of witness of being distinct.
But wait! We are just struggling to produce such a negative witness of non-equality, which in turn requires itself. A vicious cycle!
At this very point, there is no other way than resorting to the ancient cursed spell \hask{unsafeCoerce}:
\begin{code}
import Unsafe.Coerce

sa %=== sb = case testEquality sa sb of
  Just Refl -> STrue
  Nothing -> unsafeCoerce SFalse  
\end{code}
This use of \hask{unsafeCoerce} is inherently inevitable.
Fortunately, provided that \hask{TestEquality} instance is implemented soundly, this use of \hask{unsafeCoerce} is not cursed: this is just postulating an axiom that is true but there is no way to tell it to the compiler safely.
If one wants to construct evidence of type-level (non-)equality solely from the expression, we must assume some axiom and introduce it by \hask{unsafeCoerce}.
This is how library builders usually do when they implement basic (expression-level) operators to manipulate type-level values.
Such ``trust me'' axioms can be found, for example, in \hask{TestEquality TypeRep} instance in \texttt{base}, and various \hask{SEq} instances in \texttt{singletons}~\cite{singletons} package\footnote{There is another way of introducing axioms: invoking type-checker plugins.}.
Anyway, we are finally at the point of implementing working \hask{sGCD}, replacing every occurrence of \hask{(==)} with our custom \hask{(===)}:

\begin{code}
type GCD n m = GCD_ (n === 0) (m === 0) n m

type family GCD_ nEq0 mEq0 n m :: Nat where
  GCD_ 'True  _      _ m = m
  GCD_ 'False 'True  n _ = n
  GCD_ 'False 'False n m = 
    GCD_ (Mod m n === 0) 'False (Mod m n) n

sGCD :: SNat n -> SNat m -> SNat (GCD n m)
sGCD sn sm =
  case (sn %=== sZero, sm %=== sZero) of
    (STrue, _) -> sm
    (SFalse, STrue) -> sn
    (SFalse, SFalse) -> sGCD (sMod sm sn) sn
\end{code}

Finally, the compiler gets happy with all the definitions!
We can confirm that the above \hask{sGCD} works just as expected:
\begin{repl}
>>> sGCD (sNat @12) (sNat @30)
6
\end{repl}

\subsection{Summary}
When writing closed type-families, it is useful to introduce arguments \emph{witnessing} evaluation paths explicitly, interpreting clauses in a closed family as mutually exclusive alternatives.
It makes it easy to write its demoted singletonised functions.
As there are several variants of type-level equalities, we must choose appropriate one carefully; in the GCD case it is convenient to use the type-level Boolean equality which takes only reflexivity into account.

\section{Disjunctive Constraints}\label{sec:disj}
It is sporadically complained that type-classes in Haskell lack \emph{disjunction}, or \emph{logical-or}.
As for general type-classes, excluding disjunction from the language is a rational design decision for several reasons:
\begin{enumerate}
  \item Haskell adopts the open-world hypothesis for type-classes: users can add new instances, so the result of the instance resolution can differ in context.
  \label{item:openness}
  \item The semantics is not clear when multiple disjunctive clauses can be satisfied simultaneously.
  \label{item:arb-choice}
\end{enumerate}
In some cases, however, the above obstacles can be ignored, and in most cases, unrestricted disjunctions are unnecessary.
For example, when one wants to switch implementations of instances based on particular shapes of a constructor, it is customary to use \hask{{-# OVERLAPPABLE #-}} or \hask{{-# INCOHERENT #-}} pragmas.
In some cases, this makes instance resolution unpredictable and incoherent, as indicated by the name, and sometimes doesn't work well with advanced type hackery.

We propose another way to emulate disjunctive constraints, applying the constructive point of view.
First, we recall the ``meaning'' of disjunction $\varphi \vee \psi$ in BHK interpretation:
\begin{gather*}
      A: \varphi \vee \psi \iff A = \braket{i, B},
    \\
    \text{where }
    i = 0 \text{ and } B: \varphi, \text{ or }
    i = 1 \text{ and } B: \psi.
\end{gather*}
That is, a ``witness'' of $\varphi \vee \psi$ is a union of those of $\varphi$ and $\psi$.

It suggests that if constraints of interest can be recovered from \emph{witnesses}, we can take their disjunction by choosing one of such witnesses.
If such witnesses can be computed statically and deterministically, the obstacle \Cref{item:openness} is not a problem.
Further, if users can explicitly \emph{manipulate} such witnesses concretely, we can control the selection strategy of disjunctive clauses, enabling us to resolve \Cref{item:arb-choice} manually.

\subsection{Example: Field Accessors of Union Types}
OK, let us see the example.
Here, we consider the following \hask{HasFactor}-class:

\begin{code}
class HasFactor a b where
  getFactor :: b -> a
\end{code}

The intended semantics of \hask{HasFactor a b} is that ``a type \hask{b} has at least one field of type \hask{a}'', and \hask{getFactor} is a corresponding field accessor.
In short, the problem we want to tackle in this section is as follows:
\begin{quote}
  \emph{How can we give a} \hask{HasFactor a}\emph{-instance for union types}?
\end{quote}

\subsubsection{Extensible Records}
To make the situation simpler, let us introduce another player into the scene: the type \hask{Record h ks} of \emph{extensible records}.
That is, the type \hask{Record h ks} is isomorphic to a record with field labels \hask{ks :: [key]}, where each label \hask{k} in \hask{ks} is associated with a value of type \hask{h k}.
Here, we implement \hask{Record} as a heterogeneous linked-list\footnotemark:
\footnotetext{%
In practice, it is much more convenient to allow specifying labels and corresponding field type independently, i.e.\ parametrise \hask{Record} over \hask{h :: key -> val -> Type}.
And for efficiency, the underlying representation of an extensible record should be an array or vector with $O(1)$ random access.
Furthermore, it is desirable to use type-level ordered maps instead of lists to tune-up type-checking speed.
Several efficient implementations can be found on Hackage~\cite{Kinoshita:2020aa,Sterling:2020aa,Thiemann:2020aa}, and we are also planning to publish the in-house package (still in progress, though).}
\begin{code}
data Record h keys where
  EmptyRecord :: Record h '[]
  (:<) :: h k -> Record h ks -> Record h (k ': ks)
\end{code}
We also need a field accessor to retrieve information from a record:
\begin{code}
data Index k ks where
  Here  :: Index k (k ': ks)
  There :: Index k ks -> Index k (k' ': ks)

-- ...Obvious singletons and Known instances...

walkIndex :: Index k ks -> Record f ks -> f k
walkIndex Here (v :< _) = v
walkIndex (There trail) (_ :< rest) = 
  walkIndex trail rest  
\end{code}
A type \hask{Index k ks} is a \emph{witness} of the membership of a label \hask{k} in a type-level list \hask{ks}.
The function \hask{walkIndex} walks an extensible record along a given \hask{Index k ks} and retrieves a value of \hask{h k}\footnote{The same remark on efficiency as above also applies here: it is much more practical to use an index number represented as a newtyped \hask{Int} if one needs $O(1)$ random access on fields.
In such a case, one has to use \hask{unsafeCoerce} carefully to convince the compiler.}.

We can also compute \hask{Index} at type-level in an obvious way\footnote{Oh, you noticed something? Well, in \Cref{sec:plugins}, we will turn to the implementation of \hask{FindIndex} again.}, where \haskinline{(<$>)} is a type-level analogue of \hask{fmap}:
\begin{code}
type family FindIndex k ks where
  FindIndex _ '[] = 'Nothing
  FindIndex k (k ': ks) = 'Just 'Here
  FindIndex k (_ ': ks) =
    'There <$> FindIndex k ks  

type FindIndex' k ks = 
  FromJust ('Text "not found") (FindIndex k ks)
\end{code}
\hask{FindIndex'} is a variant of \hask{FindIndex}, which returns a raw membership if present, and throws a type-error otherwise.

Now that we can compute the indices of labels statically, one can implement a variant of \hask{walkIndex} with the index statically inferred:
\begin{code}
type Member k ks = Known (FindIndex' k ks)
getRecField :: forall k ks h.
  Member k ks => Record h ks -> h k
getRecField = walkIndex $ 
  demote $ sing @(FindIndex' k ks)
\end{code}
The idea is that if the concrete value of \hask{FindIndex' k ks} is \hask{Known} at the compile-time, we can use it to retrieve a field in a record.
Note that if the value of \hask{FindIndex key keys} was \hask{'Nothing}, \hask{FindIndex' key keys} reduces to a type-error.
In such a case, since the type-level language of Haskell is strict, the entire constraint \hask{Known (FindIndex' key keys)} throws a type-error during instance resolution:
\begin{repl}
>>> getRecField @Bool (EmptyRecord @Maybe)
Key `Bool' is absent in the list: '[]

>>> getRecField @Bool (['a'] :< [True, False]
                :< ([] :: [()]) :< EmptyRecord)
[True, False]
\end{repl}

Now, we can give an implementation of \hask{HasFactor} for records:
\begin{code}
instance (Member a keys, h ~ h')
      => HasFactor (h' a) (Record h keys) where
  getFactor = getRecField
\end{code}

\subsubsection{Factor of a Union}
Now we enter the situation where we want disjunctive constraints.
Consider the following type paring two records together:
\begin{code}
data RecUnion h ls rs = 
  UnionRec { recL :: Record h ls
           , recR :: Record h rs }
\end{code}
Now, how can we implement instances of \hask{HasFactor} for \hask{RecUnion}s?
Na\"{i}vely, one might at first want to write an instance like:
\begin{code}
instance (HasFactor (h a) (Record h ls)
          `Or` HasFactor (h a) (Record h rs))
  => HasFactor (h a) (RecUnion h ls rs) where ...
\end{code}
But, as noted before, there is no such things in Haskell as restricted \hask{Or}-constraint.

Fortunately, our \hask{HasFactor (h a) (Record h ls)} instance is \emph{witnessed} by a concrete type: \hask{Index a ls}!
Since its concrete value can be computed statically by \hask{FindIndex} type-family, why not combining the results of them\footnote{Here, \hask{(<|>)} is a type-level left-biased choice operator on \hask{Maybe}s.}?
\begin{code}
type UnionedIdnex k ls rs = 
  'Left <$> FindIndex k ls
  <|> 'Right <$> FindIndex k rs
type UnionedIndex' k ls rs = FromJust 
  ('Text "No label found") (UnionedIndex k ls rs)

instance Known (UnionedIndex' a ls rs)
  => HasFactor (h a) (RecUnion h ls rs) where
  getFactor (UnionRec l r) =
    case sing @(UnionedIndex' a ls rs) of
      SLeft pth -> withKnown pth $ getFactor l
      SRight pth -> withKnown pth $ getFactor r
\end{code}
Everything seems fine, but then GHC complains on call-sites of \hask{getFactor}:
\begin{repl}
• Could not deduce 
  (Known (FromJust (...) (FindIndex a ls)))...
\end{repl}
The error seems weird at first glance: we are just giving the \hask{Known} dictionary with \hask{withKnown pth}, where \hask{pth} is of type either \hask{Index a ls} or \hask{Index a rs}. Why?

The root cause of this error is that the type-checker doesn't know the following facts:
\begin{enumerate}
  \item \hask{FromJust ('Text ...)} commutes with \haskinline{(<$>)}, and
  \item If \hask{FromJust x m} reduces, then \hask{m ~ 'Just (FromJust z m)} for any \hask{z}.
\end{enumerate}
Although these two facts seem rather obvious, it needs some non-trivial axioms to infer.
Hence, if we want to convince the compiler without modifying the instance definitions, we have to augment the compiler with type-checker plugins~\cite{GHC-Team:2020aa}.
Although writing type-checker plugins is a fun, but it is not so easy to implement it \emph{correctly}. Is there any other way to avoid this obstacle?

At this point, we must notice one fact: we can still use \emph{any} value of type \hask{Index k ks} to retrieve a value of type \hask{h k} from \hask{Record h ks}.
Indeed, as \hask{ks} can have duplicated elements, there can be more than one distinct \hask{Index k ks} at the same time, e.g.\ for \hask{Index 3 '[3 ,5, 4, 3]}.
\hask{FindIndex} was just a canonical way of computing the left-most such index, if present.
To summarise, requiring \hask{Known (LookupIndex' k ks)} in \hask{getRecField} and the \hask{HasFactor}-instance for extensible records was just too much.
What we really need is a constraint demanding ``there is at least one value of type \hask{Index k ks} given'', embodied by the following class and helper functions:
\begin{code}
class Given a where
  given :: a

give :: a -> (Given a => r) -> r
\end{code}
These are excerpted from widely-used \texttt{reflection} package~\cite{Kmett:2020aa} which implements Implicit Configuration~\cite{Kiselyov:2004aa}.

With this, we can rewrite \hask{getRecField} and the corresponding \hask{HasFactor}-instance in more robust way:
\begin{code}
type Member k ks = Given (Index k ks)

getRecField
  :: Member key keys => Record h keys -> h key
getRecField = walkIndex given

-- | Serves as a default instance for @Member@.
-- Can be safely overridden by 'give' operator.
instance Known (FindIndex' k ks)
      => Given (Index k ks) where
  given = demote $ sing @(FindIndex' k ks)

instance (Given (Index a keys), h ~ h')
      => HasFactor (h' a) (Record h keys) where
  getFactor = getRecField
\end{code}

With this, we can now successfully implement \hask{HasFactor}-instance for \hask{RecUnion} as follows:

\begin{code}
-- To avoid orphan @Given@ instance
newtype IndexUnion k ls rs = WrapIdxUnion
  {unUnionIdx :: Either (Index k ls) (Index k rs)}

instance Known (UnionedIndex' k ls rs)
  => Given (IndexUnion k ls rs) where
  given = WrapIdxUnion $ demote $ 
    sing @(UnionedIndex' k ls rs)

instance Given (IndexUnion k ls rs)
      => HasFactor (h k) (RecUnion h ls rs) where
  getFactor (UnionRec l r) =
    case unUnionIdx
      $ given @(IndexUnion k ls rs) of
      Left pth -> give pth $ getFactor l
      Right pth -> give pth $ getFactor r
\end{code}

In above two implementations, \hask{Given} instances serve as a ``default instances'' to calculate witnesses.
As already shown in the implementation of \hask{getFactor} for \hask{UnionRec}, it can be overridden by \hask{give} operator without resulting in type errors complaining about overlapping instances or ``Could not deduce (Known ...)''.
This flexibility is the reason we chose to use \hask{Given} class instead of the \texttt{ImplicitParams} GHC extension providing similar functionality of dynamic scoping:  the instance shadowing in \texttt{ImplicitParams} can have unpredictable behaviour.
Such ``defaulting'' cannot be done for \hask{Known}, because type family applications such as \hask{FindIndex k ks} cannot appear at RHS of instance declarations.

Let us check it works as expected:

\begin{repl}
>>> theRec = Const "Hehe" :< Const "Foo"
      :< EmptyRecord
      :: Record (Const String) '[5,42]
>>> anotherRec = Const "Phew" :< Const "Wow"
      :< EmptyRecord
      :: Record (Const String) '[94, 5]
>>> unioned = UnionRec theRec anotherRec
>>> getFactor @(Const String 42) unioned
Const "Foo"

>>> getFactor @(Const String 94) unioned
Const "Phew"

>>> getFactor @(Const String 5) unioned
Const "Hehe"

-- Beware of reordering:
>>> getFactor @(Const String 5) 
      (UnionRec anotherRec theRec)
Const "Wow"

>>> getFactor @(Const String 999) unioned
A field of type `999' not found in either:
   Left: '[5, 42]
  Right: '[94, 5]
\end{repl}

\subsection{Summary}
We saw that we could emulate a disjunction of type constraints if the constraints in question have \emph{witnessing} type, and there is a canonical way of statically computing such witnesses at type-level.
In the above example, \hask{Member k ks} and \hask{HasFactor (h a) (RecUnion h ls rs)} are such constraints, witnessed by \hask{Index k ks} and \hask{IndexUnion k ls rs}, and such witnesses can be computed by \hask{FindIndex'} and \hask{UnionedIndex'}, respectively.
In expressing the existence of witnesses, it allows a precise and robust handling to use the \hask{Given} class.
Especially, this makes it easy to give ``fallback'' witness when canonical witness constructor can fail.
Although we treated only extensible records here, we can apply the same technique to more general settings;
we refer curious readers to \texttt{Data.Type.Path} module %
\ifbool{maingo}{%
  in the accompanying repository~\cite{demotion-repo}%
}{in the accompanying supplementary material}.

\section{Case Study: Dependently-typed Plugin System Type-Checked Dynamically}
\label{sec:plugins}

As an application of the methods developed in \Cref{sec:gcd,sec:disj}, now we look into a more involved and practical example: demoting existing type-level constraints and resolve them dynamically.

Suppose we have a program that reads a given input store and returns some outputs generated by prespecified plugins.
An input store is represented as an extensible record, and plugins are also specified in type-level.
The following signature for the main logic illustrates the idea:
\begin{code}
type SPlugins ps = SList (ps :: [Plugin])
processStore :: All (RunsOn keys) ps
  => Store keys -> SPlugins ps -> Outputs ps

class IsPlugin (p :: Plugin) where
  data PStoreType p
  data POutput p
  type Runnable p (ks :: [StoreKey]) :: Constraint
  process :: Runnable p keys
    => proxy p -> Store keys -> POutput p

class Runnable p keys => RunsOn keys p
instance Runnable p keys => RunsOn keys p
\end{code}
Thus, each plugin \hask{p} is given with a type-level constraint \hask{Runnable p keys} (and its flipped version \hask{RunsOn keys p}) to determine if it is runnable with input stores with the given keys.
A function \hask{processStore} takes an input store and (a singleton of) a list of plugin runnable on the given store and then returns the final outputs.
This works well if one specifies type variables \hask{keys} and \hask{ps} \emph{statically}.
Suppose that, however, now the situation is changed and we want to determine \hask{ps} \emph{dynamically}, depending on external configurations.
This poses two challenges:
\begin{enumerate}
  \item We have to resolve constraint \hask{All (RunsOn keys) p} at runtime, and
  \item We have to demote type-level operators as singletonised functions.
\end{enumerate}
In this section, we will see how we can apply the witness-pattern to achieve these goals safely.

\subsection{Overview of the Architecture of the Static API}\label{sec:static-api}
Before we get into the dynamic constraint resolution with witnesses, we first skim through the original static API we consider.

Our example system is designed using extensible records, which we developed in \Cref{sec:disj}.
We present the entire program architecture in \Cref{lst:plugin-arch}.
\begin{listing}[tbp]
\begin{code}
data Plugin = Doubler | Greeter
data StoreKey = IntStore | Name
              | PStore Plugin

class IsPlugin (p :: Plugin) where
  data PStoreType p
  data POutput p
  type Runnable p (ks :: [StoreKey]) :: Constraint
  process :: Runnable p keys
    => proxy p -> Store keys -> POutput p

newtype StoreEntry k =
  MkStoreEntry {storeEntry :: StoreVal k}
type Store = Record StoreEntry
type Outputs = Record POutput

type family StoreVal key where
  StoreVal 'IntStore = Int
  StoreVal 'Name = String
  StoreVal ('PStore p) = PStoreType p

type SPlugins ps = SList (ps :: [Plugin])
processStore :: All (RunsOn ks) ps
  => Store ks -> SPlugins ps -> Outputs ps
processStore _ SNil = EmptyRecord
processStore store (SCons p ps) = 
  process p store :< processStore store ps
\end{code}
\caption{Static API of a Plugin System}
\label{lst:plugin-arch}
\end{listing}
As described before, \hask{processStore} is our main routine.
It processes \hask{Store} represented as an extensible record with labels of kind \hask{StoreKey}, calculates \hask{POutput} for each plugin specified by a type-level list \hask{ps}, and bundles them into a single extensible record \hask{Outputs ps}.
\hask{StoreKey} can be either \hask{IntStore}, \hask{Name}, or \hask{PStore p} for some plugin \hask{p}.

Let's see an actual implementations for plugins \hask{Doubler} and \hask{Greeter}.
\hask{Doubler} (\Cref{lst:plugin-double}) is simple: it reads the value of \hask{IntStore}, returns the value multiplied by $2$.
\begin{listing}[tbp]
\begin{code}
instance IsPlugin 'Doubler where
  data PStoreType 'Doubler = DoubleStore
  newtype POutput 'Doubler = OutputA Int
  type Runnable 'Doubler keys = 
    Member 'IntStore keys
  process _ store = OutputA $
    2 * getRecField @'IntStore store
\end{code}
\caption{An implementation of \textt{Doubler}.}
\label{lst:plugin-double}
\end{listing}
As it requires the value associated with \hask{IntStore}, \hask{Runnable 'Doubler} demands \hask{Member 'IntStore keys}.
Hence, if there is no field with label \hask{IntStore}, it fails to type-check.

Another example, \hask{Greeter}, in \Cref{lst:plugin-greet}, is much more complicated.
\begin{listing}[tbp]
\begin{code}
type Greetable keys =
  ( Known (FindIndex ( 'PStore 'Greeter) keys)
  , Greetable_ 
    (FindIndex ( 'PStore 'Greeter) keys) keys
  )

type family Greetable_ m keys :: Constraint where
  Greetable_ ( 'Just path) _ = ()
  Greetable_ 'Nothing keys = 
    Known (FindIndex 'Name keys)

makeGreet :: PStoreType 'Greeter -> String
instance IsPlugin 'Greeter where
  type Runnable 'Greeter keys = Greetable keys
  data PStoreType 'Greeter = GreetEnv
    { greetTarget :: String
    , greetTimes :: Int, greetOwner :: String }
  newtype POutput 'Greeter = 
    GreetOutput String deriving (Show)
  process _ (store :: Store keys) =
    case sing
        @(FindIndex ('PStore 'Greeter) keys) of
     SJust idx -> withKnown idx $ GreetOutput $
       makeGreet $ getRecField 
         @('PStore 'Greeter) store
     SNothing ->
      case sing @(FindIndex 'Name keys) of
       SJust idx -> withKnown idx $ GreetOutput $
         makeGreet $ GreetEnv
          (getRecField @'IntStore store)
          1 "Greeter"
       SNothing -> GreetOutput "Hi, someone!"
\end{code}
\caption{An implementation of \texttt{Greeter}.}
\label{lst:plugin-greet}
\end{listing}
Its logic is as follows:
\begin{enumerate}
  \item If there is a field with label \hask{PStore Greeter}, generate greeting message based on it;
  \item if \hask{Name} is given, greet once to that name;
  \item otherwise, return the default greeting message.
\end{enumerate}
In short, \hask{Greeter} involves a fallback strategy on fields in the input store.
This strategy is formulated as a \hask{Greetable} constraint; first, try to inspect the existence of the label \hask{PStore Greeter}, and then fallback to the condition on the existence of \hask{PStore}.
This fallback strategy can be seen as an amalgamation of methods presented in \Cref{sec:gcd,sec:disj}, giving another way to emulate disjunction of constraints\footnote{%
In this case, however, it would suffice to require both \hask{Known} constraints in conjunction. However, in more complex cases, it saves compilation time to use such fallback strategies.}.

Let's see some examples:
\begin{repl}
>>> processStore 
      (MkStoreEntry @Name "Superman"
        :< MkStoreEntry @IntStore 42
        :< EmptyRecord)
      (sing @'[Doubler, Greeter])
OutputA 84
  :< GreetOutput "Hi, Superman, from Greeter!"
  :< EmptyRecord

>>> processStore 
      (MkStoreEntry @Name "anonymous"
        :< EmptyRecord) 
      (sing @'[ Doubler])
Key `'IntStore' is absent in: '[ 'Name]

>>> processStore 
      (MkStoreEntry @Name "Ignored"
        :< MkStoreEntry @(PStore Greeter) 
            (GreetEnv "You" 2 "me")
        :< EmptyRecord) (sing @'[ 'Greeter])
GreetOutput "Hi, Hi, You, from me!" :< EmptyRecord
\end{repl}

\subsection{Make it dynamic}
OK, let's implement a dynamic variant of \hask{processStore}.
In particular, we will make the resolution of \hask{Runnable} constraints dynamic, implementing the following function:

\begin{code}
processStoreDynamic :: Known keys
  => Store keys -> [Plugin]
  -> Either String 
    (SomeRec (RunsOn keys) POutput)

data SomeRec c f where
  MkSomeRec :: All c keys => Sing keys
    -> Record f keys -> SomeRec c f
\end{code}

One nontrivial challenge here is to resolve constraints of form \hask{Runnable} at \emph{runtime}, not \emph{compile-time}.
First, there is a well-known design pattern to allow such instance resolutions at runtime: the \emph{deferrable constraint} pattern:
\begin{code}
class Deferrable p where
  deferEither :: proxy p -> (p => r) 
              -> Either String r
\end{code}
This is provided in \hask{Data.Constraint.Deferrable} module from \texttt{constraints} package~\cite{Kmett:2020ab}, and it was first proposed by Dimitrios Vytiniotis.
For example, assume the following singletonised version of \hask{FindIndex}:
\begin{code}
sFindIndex :: Sing key -> SList keys
      -> SMaybe (FindIndex key keys)
\end{code}
Then we can implement \hask{Deferrable} for \hask{Member k ks}:
\begin{code}
instance (Known k, Known ks)
  => Deferrable (Member k ks) where
  deferEither _ = 
    case sFindIndex (sing @k) (sing @ks) of
      SJust n -> give (demote n) $ Right act
      SNothing -> Left "Not found"
\end{code}
This illustrates another advantage of using \hask{Given} instead of \hask{Known} in \hask{Member}: if we had formalised \hask{Member} in terms of \hask{Known}, we cannot implement \hask{Deferrable} for \hask{Member} because it includes irreducible type-family constructor, i.e.\ \hask{FindIndex k ks}.
Unfortunately, for the same reason, we cannot write direct \hask{Deferrable}-instances for \hask{Runnable p keys} in general, as \hask{Runnable} is a \emph{type family} and instance declarations cannot include type family application in their header.

Instead, we provide the dedicated class \hask{DynamicPlugin} of plugins which allow the deferral of corresponding \hask{Runnable}s:
\begin{code}
class IsPlugin p => DynamicPlugin p where
  deferDynamicPlugin
    :: Known keys
    => pxy p -> Proxy keys
    -> (Runnable p keys => r) -> Either String r
\end{code}
So it remains to implement \hask{IsPlugin} instances for each plugin.
Fortunately, all the \hask{Runnable} definitions defined so far can be resolved with singleton manipulation.

Let's see how to resolve \hask{Runnable 'Doubler keys} dynamically.
Recall its definition:
\begin{code}
type Runnable 'Doubler ks = Member 'IntStore ks
\end{code}
By definition, \hask{Member 'IntStore keys ~ Given (Index 'IntStore keys)} and the default implementation is resolved with the superclass constraint \hask{Known (FindIndex'  'IntStore keys)}.
we can implement \hask{DynamicPlugin 'Doubler} as follows:
\begin{code}
instance DynamicPlugin 'Doubler where
  deferDynamicPlugin _ (_ :: Proxy keys) =
    deferEither_ @(Member SIntStore keys)
\end{code}
We can likewise implement the instance for \hask{Greeter}:
\begin{code}
instance DynamicPlugin 'Greeter where
 deferDynamicPlugin _ (_ :: Proxy keys) act =
  case sFindIndex (SPStore SGreeter) keys of
   SJust pth -> withKnown pth $ Right act
   SNothing -> withKnown (sFindIndex SName keys)
            $ Right act
  where keys = sing @keys
\end{code}

Hence, it remains to implement \hask{sFindIndex}.
Recall our current implementation of \hask{FindIndex}:
\begin{code}
type family FindIndex k ks where
  FindIndex _ '[] = 'Nothing
  FindIndex k (k ': ks) = 'Just 'Here
  FindIndex k (_ ': ks) =
    'There <$> FindIndex k ks  
\end{code}
Readers might notice that modification is needed to distinguish cluases by \emph{witnesses}, as we did in \Cref{sec:gcd}:
\begin{code}
type FindIndex :: forall k ks -> Maybe (Index k ks)
type family FindIndex k ks where
  FindIndex _ '[] = 'Nothing
  FindIndex k (k' ': ks) = 
    FindIndexAux (k === k') k ks

type FindIndexAux
  :: forall k'. Bool -> forall k ks
  -> Maybe (Index k (k' ': ks))
type family FindIndexAux eql k rest where
  FindIndexAux 'True _ _ = 'Just 'Here
  FindIndexAux 'False k ks = 
    'There <$> FindIndex k ks
\end{code}
OK, now we can implement \hask{sFindIndex}:
\begin{code}
sFindIndex :: TestEquality (Sing @a)
  => Sing (k :: a) -> SList keys
  -> SMaybe (FindIndex k keys)
sFindIndex _ SNil = SNothing
sFindIndex k (SCons k ks') = case k %=== k' of
  STrue -> SJust SHere
  SFalse -> SThere %<$> sFindIndex k ks
\end{code}
...Well, not quite:
\begin{repl}
• Could not deduce: 
    FindIndexAux k x xs 'True ~ 'Just a0
  Expected type: SMaybe (FindIndex k keys1)
    Actual type: SMaybe ('Just a0)
  The type variable ‘a0’ is ambiguous
• In the expression: SJust SHere
  In a case alternative: STrue -> SJust SHere
\end{repl}
We expected \hask{a} to be \hask{Here :: Index k (k ': ks)}, but here it becomes ambiguous type variable. Why?
In the \hask{'True}-branch of \hask{FindIndexAux}, it returns \hask{Here}.
Recall that \hask{Here} has the following type:
\begin{code}
  Here :: k ~ k' => Index k (k' ': ks)
\end{code}
This tells us the reason why we got stuck: in \hask{STrue}-branch, GHC knows that \hask{(k === x) ~ 'True}, but GHC cannot infer \hask{k ~ x} from it!

To avoid such information loss, it is convenient to pack all extensionally equivalent constraints.
Recall we have three type-level (homogeneous) equalities:
\begin{enumerate}
\item \hask{a ~ b}, the built-in equality constraint,
\item \hask{a == b}, a type-level boolean predicate that plays well with compound types but lacks automatic reflexivity, and
\item \hask{a === b}, a type-level boolean predicate which takes only the reflexivity into account.
\end{enumerate}
These three equalities can play their roles case-by-case, although their extensions must coincide.
In addition, GHC cannot tell that latter two equalities are symmetric.
So, it is useful to pack all these into a single witness, as follows:
\begin{code}
data Equality a b where
  Equal :: ((a == b) ~ 'True, (b == a) ~ 'True,
    (a === b) ~ 'True, (b === a) ~ 'True, a ~ b)
    => Equality a b
  NonEqual
    :: ((a === b) ~ 'False, (b === a) ~ 'False, 
        (a == b) ~ 'False, (b == a) ~ 'False)
    => Equality a b
\end{code}
As expressed by the constructor names, here \hask{Equal} witnesses the equality of given two types, and \hask{NonEqual} witnesses \emph{non}-equality.
In this way, \hask{Equality} packages both \emph{positive} (equal) and \emph{negative} (non-equal) witnesses.

One might wonder why we didn't mention equality \emph{constraint} \hask{a ~ b} in \hask{NonEqual}-case.
This is because GHC can infer that \hask{a :~: b} is not inhabited from \hask{(a === b) ~ 'False}:

\begin{code}
{-# LANGUAGE EmptyCase, LambdaCase #-}
import Data.Void
fromFalseEq
  :: (a === b) ~ 'False => a :~: b -> Void
fromFalseEq = \case {}
\end{code}
To allow equality test with \hask{Equality} witnesses, we use the following class:
\begin{code}
class SEqual k where
  (%~) :: Sing (a :: k) -> Sing b -> Equality a b
\end{code}
Some reader might realise that this looks similar to \hask{SDecide} class in \texttt{singletons} package~\cite{singletons}.
The difference lies in the non-equal case: \haskinline{(
in \hask{SDecide} returns \hask{a :~: b -> Void} when non-equal.
This design decision works well if one only do with \hask{(~)}; however, we cannot derive \hask{(a == b) ~ 'False} or \hask{(a === b) ~ 'False} by the very same reason why we cannot use \haskinline{(
in the definition of \hask{sFindIndex}.

Anyway, since those three equalities coincides extensionally, we can derive instance definitions of \hask{SEqual} from either \hask{TestEquality} (from \texttt{base}) or \hask{SDecide}.
One can of course directly implement \hask{SEqual} by pattern matching.
For example:
\begin{code}
instance SEqual Plugin where
  SDoubler %~ SDoubler = Equal
  SDoubler %~ SGreeter = NonEqual
  SGreeter %~ SGreeter = Equal
  SGreeter %~ SDoubler = NonEqual
\end{code}
For \hask{StoreKey}, however, we need some hacks as it involves compound constructors:
\begin{code}
instance SEqual StoreKey where
  SIntStore %~ SIntStore = Equal
  SIntStore %~ SName = NonEqual
  SIntStore %~ SPStore {} = NonEqual
  -- ... Similar for SName %~ x ...
  SPStore p %~ SPStore q = case p %~ q of
      Equal -> Equal
      NonEqual -> nestNonEqual
  SPStore {} %~ SIntStore = NonEqual
  SPStore {} %~ SName = NonEqual

nestNonEqual :: (a == b) ~ 'False => Equality a b
nestNonEqual = unsafeCoerce $ NonEqual @0 @1
\end{code}
The reason we cannot use \hask{NonEqual} in the \hask{SPStore}-case is that, as state before, \hask{(===)} won't inspect inside compound types.
On the other hand, since \hask{(==)} from \texttt{base} can correctly handle compound types, we can call combinator \hask{nestNonEqual} here.
By constraining with \hask{(a == b) ~ 'False}, it tries to reject illegal usages as much as possible\footnote{The use of \hask{unsafeCoerce} here is memory-safe, even without any constraints, because coercion arguments has no runtime representation.}.

With \hask{SEqual} above, we can now implement \hask{sFindIndex}:
\begin{code}
sFindIndex :: SEqual a => Sing (k :: a)
  -> SList keys -> SMaybe (FindIndex k keys)
sFindIndex _ SNil = SNothing
sFindIndex k (SCons k' ks) =
  case k %~ k' of
    Equal -> SJust SHere
    NonEqual -> SThere %<$> sFindIndex k ks
\end{code}

As this filled the last gap, we can now implement our last goal \hask{processStoreDynamic} as listed in \Cref{lst:proc-dyn}.
\begin{listing}[tbhp]
\begin{code}
data PluginsOn keys where
  MkSomePlugins :: All (RunsOn keys) ps
    => SPlugins ps -> PluginsOn keys

toSomeDyns :: forall keys. Known keys
  => [Plugin] -> Either String (PluginsOn keys)
toSomeDyns [] = pure $ MkSomePlugins SNil
toSomeDyns (p : rest) = do
  MkSomePlugins ps <- toSomeDyns @keys rest
  withPromoted p $ \case
    SDoubler ->
      deferDynamicPlugin
        (Proxy @ 'Doubler) (Proxy @keys)
        (MkSomePlugins $ SCons SDoubler ps)
    SGreeter ->
      deferDynamicPlugin
        (Proxy @ 'Greeter) (Proxy @keys)
        (MkSomePlugins $ SCons SGreeter ps)

processStoreDynamic :: forall keys. Known keys
  => Store keys -> [Plugin]
  -> Either String (SomeRec (RunsOn keys) POutput)
processStoreDynamic store ps = do
  MkSomePlugins (sps :: SPlugins ps) 
    <- toSomeDyns @keys ps
  pure $ MkSomeRec sps $ processStore store sps
\end{code}
\caption{The implementation of \texttt{processStoreDynamic}}
\label{lst:proc-dyn}
\end{listing}

Now, we can test \hask{processStoreDynamic}:
\begin{repl}
>>> processStoreDynamic
      (MkStoreEntry @Name "Superman"
        :< MkStoreEntry @IntStore 42
        :< EmptyRecord)
      [Doubler, Greeter] -- it is a value!
Right (OutputA 84
  :< GreetOutput "Hi, Superman, from Greeter!"
  :< EmptyRecord)

>>> processStoreDynamic
      (MkStoreEntry @Name "anonymous"
        :< EmptyRecord) 
      [Doubler]
Left "Doubler requries IntStore key"
\end{repl}
\subsection{Summary}
We discussed the design of a statically-typed plugin system with dynamic instance resolution at runtime.
This was achieved by combining witness-aware constraint handling and the Deferrable class.
We also discussed the design of the equality witness that can treat three distinct type-level equalities.

\section{Conclusions}
\label{sec:concl}
We demonstrated how we could use the constructive point of view, paying attention to \emph{witnesses}, as a useful guiding principle in designing embedded dependent type-systems in Haskell.
As a concrete example, we have demonstrated:
\begin{enumerate}
  \item Type-level arguments \emph{witnessing} evaluation paths in a type-family enables us to safely write corresponding singletonised function much easier afterwards.
  \item Disjunctions of type constraints can be emulated if it is recoverable from \emph{witnesses} that is statically computable at type-level.
  \item Combining \emph{witnesses} with \hask{Deferrable} class, we can implement a dependently-typed plugin system, which can be dynamically type-checked at \emph{runtime}.
  \item It is convenient to provide a unified witness type for extensionally equivalent, but not definitionally equivalent constraints; we took type-level equalities as an example.
\end{enumerate}
To summarise, what we demonstrated in this paper is not a single \emph{method}, but an insightful way of thinking when one designs dependently typed programs.

\subsection{Related and future works}
The examples of this paper incorporates many existing works on dependent types in Haskell: those include type-families~\cite{Kiselyov:2010aa}, data-type promotion~\cite{Yorgey:2012}, singletons~\cite{Eisenberg:2012}, Implicit Configuration~\cite{Kiselyov:2004aa}, to name a few.
The Hasochism paper~\cite{10.1145/2503778.2503786} contains many examples of dependently-typed programming and obstacles in Haskel.
We demonstrated how useful the constructive point of view is when we write dependently-typed codes incorporating these prominent methods.
There is a minor difference in the direction of interest, though.
In dependently-typed programming in Haskell, the method to \emph{promote} functions and data-types is widely discussed~\cite{Yorgey:2012,Eisenberg:2012,10.1145/2503778.2503786}.
On the other hand, many examples in the present paper arise when one tries to \emph{demote} type-level hacks down to the expression-level.
Such needs of demotion arise when one wants to resolve dependently-typed constraints at runtime, as we saw in \Cref{sec:plugins}.
The demotion-based approach has another advantage compared to the promotion-based one: the behaviours of macros for deriving definitions are much predictable.
For example, in \texttt{singletons}~\cite{singletons} package, there are plenty of macro-generated type-families with names with seemingly random suffixes.
This makes, for example, writing type-checker plugins to work with them rather tedious.

We have used user-defined witnesses carrying instance information, such as singletons and equality witness.
If once coherent explicit dictionary applications~\cite{Winant:2018wu} get implemented in GHC, we will be able to directly treat instance dictionaries as another kind of witnesses.

Since the contribution of this paper is a general point of view, there is much room for exploration of synergy with other methods with dependent types.
For example, the method of Ghosts of Departed Proofs~\cite{Noonan:2018aa} shares the witness-aware spirit with the present paper.
It suggests aggressive uses of phantom types to achieve various levels of type-safety.
It can be interesting to explore the synergy of such methods with our examples.
For example, we can use the \hask{Reified} class with default instances instead of \hask{Given} to switch selection strategies based on phantom type parameters.

The compilation performance matters much when one tries to apply involved type-level hacks in the industry and there are many possibilities for exploration in this direction.
For example, suppose we promote container types, such as trees, and provide basic construction as a type-function, rather than data-constructors.
Then, as the current GHC doesn't come with the ability to inline type-level terms, it can take so much time to normalise such type-level constructions when they appear repeatedly.
In such cases, the compiler plugins to expand type families at compile time can help to improve the compilation time; but this is only a partial workaround and more investigations must be taken.

\begin{acks}
\ifbool{maingo}{%
Many of techniques presented in this paper has been found while developing products at DeepFlow, Inc.
I thank my employer and colleagues for fruitful discussions and their challenging spirit which allows me to use type-level features so agressively.
}{%
Techniques presented in this paper were found while developing software at my company.
I thank my colleagues for fruitful discussions.
}
\end{acks}

\bibliographystyle{ACM-Reference-Format}
\bibliography{references.bib}

\appendix


\end{document}